\documentclass[aps,prb,reprint,longbibliography,superscriptaddress]{revtex4-2}
\usepackage{mathrsfs}
\usepackage{amsmath}
\usepackage{amsfonts}
\usepackage{amssymb}
\usepackage{amsthm}
\usepackage{graphicx}
\usepackage{natbib}
\usepackage{color}
\usepackage{hyperref}
\usepackage{bm}
\usepackage[caption=false]{subfig}
\usepackage{verbatim}
\usepackage{ulem}
\usepackage[english]{babel} 

\usepackage{calrsfs}
\DeclareMathAlphabet{\pazocal}{OMS}{zplm}{m}{n}

\providecommand{\U}[1]{\protect\rule{.1in}{.1in}}

\begin{document}
\title{The Physics of Superconductor--Ferromagnet Hybrid Structures}

\author{Alexander A. Golubov}
\email{a.golubov1960@gmail.com}
\affiliation{Moscow Institute of Physics and Technology, Dolgoprudny, 141700 Moscow region, Russia}
\affiliation{HSE University, 101000 Moscow, Russia}

\author{Sergey V. Bakurskiy}
	\affiliation{Skobeltsyn Institute of Nuclear Physics, Lomonosov Moscow State University, 119991 Moscow, Russia}

\author{Tairzhan Karabassov}
\affiliation{Moscow Institute of Physics and Technology, Dolgoprudny, 141700 Moscow region, Russia}
\affiliation{HSE University, 101000 Moscow, Russia}

\author{A.S. Vasenko}
\affiliation{HSE University, 101000 Moscow, Russia}

\author{Mikhail Yu. Kupriyanov}
\affiliation{Moscow Institute of Physics and Technology, Dolgoprudny, 141700 Moscow region, Russia}
\affiliation{Skobeltsyn Institute of Nuclear Physics, Lomonosov Moscow State University, 119991 Moscow, Russia}
\author{Anatolie S. Sidorenko}
\affiliation{Technical University of Moldova, 2004 Chisinau, Moldova}

\begin{abstract}
In this review, we summarize the foundations underlying a variety of phenomena in superconductor--ferromagnet hybrid structures, with a focus on recent advances in several key areas. These include: (i) the fundamental understanding of proximity effects in superconductor--ferromagnet based systems; (ii) spin-valve effects in superconductor--ferromagnet and superconductor--ferromagnet--superconductor Josephson junctions; and (iii) the design and realization of superconducting memory elements, particularly in hybrid Josephson junctions. We also discuss the experimental progress in fabricating and characterizing spin-valve structures.
\end{abstract}

\maketitle

\newpage


\section{Introduction}

The interplay between superconductivity and ferromagnetism in hybrid superconductor–ferromagnet (SF) structures continues to be a subject of intense theoretical and experimental interest. While these two phenomena are antagonistic in nature, superconductivity favoring spin-singlet pairing and ferromagnetism promoting spin polarization, their combination in mesoscopic systems leads to a wide range of nontrivial and technologically promising effects. Among these are oscillations of the critical temperature and critical current due to the  Fulde-Ferrell-Larkin-Ovchinnikov (FFLO)~\cite{Larkin1964,Fulde1964} - like behavior \cite{RevModPhys.77.935}, $\pi$-junction formation \cite{1128994620010312,Kontos2002}, and the emergence of spin-triplet pairing states in the presence of magnetic inhomogeneity \cite{RevModPhys.77.1321}. 
Understanding these effects is not only of fundamental importance but also essential for the development of superconducting spintronics and quantum memory elements.

Recent progress has demonstrated that SF-based Josephson junctions can function as controllable phase elements and memory cells in  superconducting digital electronic and spintronic circuits \cite{Linder2015, Birge2024, yang2021boosting,cai2023superconductor,Mitrovic2025}. Novel designs such as SIsFS  junctions (I is an insulator and s is a thin superconducting
layer) \cite{Bakurskiy1, Nevirkovets1} allow tuning between 0 and $\pi$ states with large $I_c R_n$ products, making them attractive for cryogenic memory applications. Comprehensive theoretical frameworks based on the Usadel equations within the quasiclassical Green’s functions formalism have successfully described much of the proximity and spin transport physics in these systems \cite{RevModPhys.77.935, RevModPhys.77.1321, RevModPhys.76.411, Eschrig2011,Bergeret2018}

The theoretical foundation for the $\pi$ state was laid by Bulaevskii et al. \cite{106323119770405}, who predicted its existence in junctions with magnetic impurities and its manifestation as spontaneous currents in superconducting rings. The experimental breakthrough came with the work of Ryazanov et al., which demonstrated the first supercurrents and the reversible 0-$\pi$ transition in SFS junctions \cite{S092145340001476320000101, 1128994620010312}, later solidified by direct phase measurements \cite{ISI:000173213100011}. This state is fundamentally characterized by a ground-state phase difference of $\pi$, corresponding to a negative critical current $I_C$ in the sinusoidal Current-Phase Relation (CPR) \cite{RevModPhys.76.411}. 
This discovery served as a catalyst for the field, driving further experimental work that uncovered a host of new phenomena in junctions incorporating various ferromagnetic materials \cite{8612398820011002, 928412920030110, S092145340401343720050101, 2808053420080101, 893023120060519, 918380320061001, Blum2002, Sellier2003, Sellier2004, Weides2006_1, Weides2006_2, robinson2006critical, robinson2007zero, Pfeiffer2008, Weides2009}. 

Physics of $0$- to $\pi $ crossover in various types of Josephson SFS junctions has been studied theoretically in different regimes by Buzdin \textit{et al.} \cite{ISI:A1982PC10900006,ISI:A1992HM98200020,vujicic1991the-oscillation724334}, Buzdin and Kupriyanov  \cite{ISI:A1991FT22600011}, Radovi\'{c} \textit{et al.}
\cite{Radovic1991}, Proki\'{c} \textit{et al.} \cite{Prokic1999}, 
Dobrosavljevi\'{c}--Gruji\'{c} \textit{et al.} \cite{Dobrosav2000}, Fogelstr\"{o}m
\cite{Fogelstr2000}, Barash and Bobkova \cite{ISI:000174980300108}, Kuli\'{c} \cite{Kulic2001}, Krivodruchko and Koshina \cite{koshina2000spin}, Fominov \textit{et al.}
\cite{Fominov2002}, Chtchelkatchev \textit{et al.}
\cite{chtchelkatchev2001pi}, Bergeret \textit{et al.}\cite{ISI:000168525900043,707903420011001}, 
Golubov \textit{et al.} \cite{729048420020225,golubov2002nonsin}, Buzdin and Baladie \cite{ISI:000183378800084} and Karminskaya \textit{et al.} \cite{Karminskaya2009}. 
Recently, Birge and Satchell explored the materials landscape for implementing Josephson $\pi$ junctions \cite{Birge2024} . In
addition to $\pi $-transitions, new intriguing predictions have been made for complex Current - Phase Relations (CPR) in SFS junctions.

Further, the generation of the long-range triplet order parameter was predicted in structures with inhomogeneous magnetization or with non-collinear orientations of magnetization in different F-layers by Bergeret \textit{et al.}  \cite{Bergeret2001} and Kadigrobov \textit{et al.} \cite{Kadigrobov2001}
(see the review \cite{RevModPhys.77.1321} and references therein). 

Variety of systems exhibiting $\pi $-states includes planar SFS proximity effect structures, tunnel junctions with magnetic insulator or magnetically  active interfaces and structures with the barriers containing more than one magnetic layer. Josephson effect in the junctions between unconventional superconductors with different types of magnetic barriers was studied
theoretically by Tanaka \textit{et al.} \cite{tanaka1999josephson, yokoyama2007theory}.

In this review, we summarize the theoretical foundations governing the key physical phenomena in SF) hybrid systems, with particular emphasis on recent advances in several directions. First, we examine the superconducting proximity effect in SF structures, highlighting the penetration of superconducting correlations into the ferromagnetic region. 
Special attention is given to the oscillatory and damped behavior of the superconducting order parameter inside the ferromagnet, which underpins many of the emergent phenomena in SF systems.
This process leads to distinctive spectral and transport features arising from the competition between superconducting pairing and spin polarization. 

A central phenomenon discussed is the spatially oscillatory and damped behavior of the superconducting pair amplitude inside the ferromagnet, a direct consequence of the exchange field-induced phase shifting of the electron and hole components of Cooper pairs. This oscillatory proximity effect forms the basis for various emergent phenomena in SF structures, such as the $0$--$\pi$ transitions in Josephson junctions, nonmonotonic and reentrant critical temperature behavior in SF bilayers and trilayers, and the generation of odd-frequency triplet correlations under conditions of spin mixing and rotation at interfaces.

We highlight recent experimental progress in realizing SF-based spin-valve structures, especially those employing homogeneous metallic ferromagnets with collinear magnetization configurations. These systems are particularly amenable to theoretical treatment and have served as effective testbeds for validating predictions related to the proximity effect, spin-triplet pairing, and magnetization-controlled Josephson switching.

Finally, we discuss the spin-valve effect in SFS and SIsFS Josephson junctions, where the critical current can be modulated by controlling the relative orientation of magnetizations in ferromagnetic layers. This tunability is of particular interest for cryogenic memory applications, where superconducting elements must reliably store and switch information. In this context, we review the design principles and theoretical modeling of Josephson memory cells based on the SIsF junction architecture, which allow for both 0 and $\pi$ phase states and non-volatile memory behavior.

Throughout this review, we focus on SF-based hybrid structures where the ferromagnetic layers are assumed to be uniform and metallic, and the magnetizations are aligned either parallel or antiparallel. This restriction simplifies the theoretical framework while still capturing the essential physics relevant to practical devices.

\section{Proximity Effect}

\subsection{Basic Concepts}

Let us consider a bilayer composed of a semi-infinite superconductor (S) and a ferromagnet (F), separated by an interface of arbitrary transparency. The superconducting correlations induced in a ferromagnet differ qualitatively from those in superconductor–normal metal (SN) proximity systems. Generally, the proximity effect refers to the penetration of Cooper pair amplitudes into a non-superconducting material, mediated by Andreev reflection. In this process, an electron and a hole with opposite spins and momenta are correlated, giving rise to a superconducting pairing amplitude within the adjacent material.

In an SN bilayer, the proximity-induced correlations decay exponentially into the N-metal due to dephasing between the electron and hole wave functions. However, in an SF bilayer, the correlated electron–hole pairs—possessing opposite spins—experience the internal exchange field of the ferromagnet. This interaction leads to an energy shift between the quasiparticles, generating a nonzero center-of-mass momentum 
$Q$ for the Cooper pairs~\cite{565636119970601}.

As a consequence, the superconducting pair amplitude in the ferromagnetic region oscillates spatially as 
$\cos(Qx)$. This sign-changing behavior is equivalent to periodic 
$\pi$ phase shifts in the F region, and it resembles the spatially modulated (FFLO) state originally proposed for magnetic superconductors~\cite{Larkin1964,Fulde1964}.

The amplitude of these oscillations decays with distance from the SF interface. The decay behavior depends significantly on the transport regime of the F-metal. In the diffusive (dirty) limit, the decay length at zero temperature coincides with the oscillation period~\cite{Radovic1991,ISI:A1991FT22600011}. In contrast, in the clean limit, the decay length becomes formally infinite at $T = 0$ and is limited in practice by elastic impurity scattering~\cite{Bergeret2002} or spin–orbit scattering~\cite{565636119970601}. Consequently, the decay length typically exceeds the oscillation period, making spatial oscillations more readily observable in clean or quasi-clean systems.

Such oscillatory behavior has been experimentally confirmed, including multi-period FFLO-type oscillations of the superconducting critical temperature in nearly clean SF bilayers~\cite{sidorenko2003oscillations,sidorenko2017reentrance}. The crossover between clean and diffusive regimes remains a central theoretical challenge in accurately modeling SFS structures and understanding their transport and spectral properties~\cite{707903420011001,918380320061001,2301897220060701,1071066920090201,1232197120111001}.

\subsection{Complex Coherence Length}

In a structure composed of superconducting and ferromagnetic films, the complex coherence length, $\xi$, is a key material parameter that characterizes the proximity effect. It defines both the spatial decay and the oscillation period of superconducting correlations penetrating into the ferromagnetic (F) layer.
In the dirty limit, the coherence length can be derived from the solution of the Usadel equations \cite{Usadel}. It takes the form:
\begin{equation}
\frac{1}{\xi }=\frac{1}{\xi _{F}}\sqrt{\frac{\omega + iH}{\pi T_{c}}},\quad
\xi _{1,2}=\xi _{F}\sqrt{\frac{\pi T_{c}}{(H^{2}+\omega ^{2})^{1/2}\pm
\omega }}
\label{xi}
\end{equation}%
where $\xi _{F}^{2}=D_{F}/2\pi T_{c},$ and $D_{F}=\ell _{F}v_{F}/3$ is the diffusion coefficient, $\omega = \pi T (2n+1)$ is the Matsubara frequency, $T_{c}$ is the critical temperature of the S part of proximity SF system, $\xi _{1,2}$ are the real and imaginary parts of the coherence length, respectively, and $H$ is the the exchange energy of the ferromagnet. The expression (\ref{xi}) is valid in the most practical regime, when the electronic mean free path $\ell$ is the smallest characteristic scale in the problem compared to
 $\sqrt{D_{F}/2 H}$, $\sqrt{D_{F}/2 \pi T_{c}}$ and the thicknesses of the S, $d_S$, and F, $d_F$, layers. In this case, strong elastic scattering leads to averaging over the electronic trajectories. 

In the clean limit and in the case of a semi-infinite F layer the expression for the decay length follows from the Eilenberger equations \cite{7207646919680201}.  

\begin{equation}
\frac{1}{\xi }=\frac{1}{\xi _{\omega}} + \frac{1}{\ell }
+ i\frac{1}{\xi _{H}},
\label{xicL}
\end{equation}
where $\xi_{\omega}=v_F /2 \omega$, $\xi_H = v_F/2H$ and $v_F$ is the Fermi velocity in the ferromagnet.

The expression for the coherence length given in Eq.~(\ref{xicL}) is valid in the limit of a thick ferromagnetic layer (F), where $d\gg l,\xi_{\omega}$. For the finite thickness of the F layer the boundary condition requires that the spatial derivative of the superconducting order parameter vanishes at the outer surface of the F layer. This constraint leads to the selection of only those order parameter configurations that exhibit an extremum at the boundary. As a result, phase synchronization effects arise, making both the real and imaginary parts of the complex coherence length, $\xi$, dependent on the F-layer thickness 
$d_F$. A detailed analysis of this crossover behavior, based on the linearized Eilenberger equations \cite{7207646919680201}, has been carried out by Pugach \textit{et al.} \cite{1232197120111001}.

In typical experimental conditions \cite{893023120060519}, the relationship between the electronic mean free path $l$
and the coherence lengths 
$\xi_1$, $\xi_2$ more closely corresponds to the dirty limit. Under such circumstances, one might expect $\xi_1 = \xi_2$.  However, experimental studies \cite{Blum2002,Weides2009} have demonstrated that a noticeable difference exists between $\xi_1$ and $\xi_2$, indicating the presence of additional scattering mechanisms or nontrivial proximity effects in realistic systems.

Several possible explanations have been proposed to account for the observed difference between the decay and oscillation lengths, $\xi_1$ and $\xi_2$, respectively.

One prominent mechanism is the presence of strong \textit{paramagnetic spin-flip scattering} in the ferromagnetic (F) layer. This process is intrinsic to ferromagnetic materials—especially dilute or weak ferromagnetic alloys such as Cu$_x$Ni$_{1-x}$ with $x \sim 0.5$, which are often used in S/F/S junction experiments. Such alloys exhibit rapid suppression of the Josephson critical current as a function of the F-layer thickness, which is indicative of strong spin-dependent scattering.

In addition to spin-flip scattering, \textit{spin-orbit scattering} and other pair-breaking mechanisms also play a significant role in modifying the superconducting proximity effect within the F layer. These interactions contribute to an effective suppression and decoherence of the spin-singlet Cooper pairs, thereby altering the characteristic lengths.

In the theoretical work by Faure \textit{et al.}~\cite{Faure2006}, it was shown that in the diffusive limit and in the presence of, respectively, spin-flip and spin-orbit scattering the complex coherence length $\xi = \xi_1 + i \xi_2$ is given by the following expressions:

\begin{equation}
\frac{\xi _{1}}{\xi _{2}}=\frac{\sqrt{\sqrt{1+\alpha ^{2}}-\alpha }}{\sqrt{%
\sqrt{1+\alpha ^{2}}+\alpha }},~
\frac{\xi _{1}}{\xi _{2}}= \sqrt{1-\alpha_{so}^2},
\label{xiR}
\end{equation}%
where  $\alpha =(H \tau _{m})^{-1},$ $\alpha_{so}=(H \tau _{so})^{-1}$,  $\tau _{m}$ and $\tau _{so}$ are
the spin-flip and spin-orbit scattering times, respectively. 
In Ref. \cite{Karabassov2022} spin-dependent scattering have been taken into account in calculations of the density of states in SF bilayers.

This relation shows that the ratio $\xi_1 / \xi_2$ deviates from unity as spin-dependent scattering processes become significant, explaining the experimentally observed distinction between the decay and oscillation lengths in ferromagnetic layers~\cite{Blum2002,Weides2009}.

The next source of the difference between $\xi_1$ and $\xi_2$ can be associated with the presence of domain walls in the F film. 
To support this assertion, Bakurskiy \textit{et al.}~\cite{edsgcl.43018830220150101} analyzed a multilayer SIFS structure composed of a superconducting electrode (S), an insulator (I), and a ferromagnet–superconductor (FS) bilayer as the upper electrode.
The F film was assumed to exhibit a domain structure in which adjacent domains possess antiparallel magnetization vectors and are separated by atomically sharp walls oriented perpendicular to the SF interfaces.

It was shown that the ratio
\begin{equation}
\frac{\xi _{1}}{\xi _{2}} = \frac{\sqrt{\gamma _{BW}^{2}h^{2} - 1}}{\sqrt{(
\gamma _{BW}\Omega + 1) ^{2} + \gamma _{BW}^{2}h^{2} - 1} + \Omega/ (\gamma
_{BW} + 1)}
\label{ratio}
\end{equation}
monotonically increases from zero at $\gamma_{BW} h = 1$ to the corresponding  value for a single-domain SIFS junction in the limit $\gamma_{BW} \rightarrow \infty$.
Here $\gamma_{BW} = \gamma_B W $/ 2, $h = H / \pi T_C$, $\Omega = \omega / \pi T_C$, and $W$ is the domain width. The suppression parameter $\gamma_B = R_B A_B / (\rho_F \xi_F)$, where $R_B$ and $A_B$ are the resistance and area of the domain-wall interface, respectively, and $\rho_F$, $\xi_F$ are the resistivity and coherence length of the F layer.

From Eqs.~(\ref{xicL}) and (\ref{ratio}), it follows that the characteristic length scale of the oscillations of superconducting correlations in the F layer can vary over a wide range, whereas the damping scale is fundamentally limited by the exchange energy $H$. The latter typically exceeds $\pi T_{c}$ even in dilute ferromagnets. As a consequence, the characteristic decay length $\xi_{1}$ of superconducting correlations in ferromagnets does not exceed several nanometers. This presents a significant technological challenge for the fabrication of SFS structures that meet the stringent requirements imposed on superconducting memory devices. A promising approach to address this issue involves replacing the ferromagnetic layer with an engineered artificial material: a multilayer stack consisting of alternating thin layers of normal (N) and ferromagnetic (F) metals. In such FN bilayers, electrons spend part of their time in the N regions, where spin ordering is absent. As a result, the electrons experience an effective exchange energy that is averaged over the total thickness of the FN bilayer. This effective exchange energy is reduced compared to the exchange energy in the pure ferromagnetic layer \cite{Bergeret2005,1291265520120801}.

Small exchange fields are crucial for realization of 0-$\pi$ transitions in SF structures because they reduce magnetic disorder and suppress interface roughness effects, preventing the smearing of the transition that would occur with stronger exchange fields due to increased inhomogeneity and stray fields. Vasenko et al. \cite{Vasenko2015} proposed methods to measure weak spin-splitting fields using the subgap conductance of SFN hybrids. This approach leverages the sensitivity of Andreev reflection processes to spin polarization, providing a tool to probe subtle magnetic effects in proximity-coupled systems. The theoretical work by Ozaeta et al. \cite{Ozaeta2012}  demonstrated that small exchange fields can significantly enhance the subgap conductance and Andreev current in such structures, offering a pathway to optimize spin-dependent transport.

Furthermore, in structures where the magnetization vectors $\bm{M}$ of the ferromagnetic layers are collinear, a transition in the mutual orientation of $\bm{M}$ between adjacent layers - from parallel (ferromagnetic) to antiparallel (antiferromagnetic) - can significantly reduce the effective exchange energy $H_{\mathrm{eff}}$, potentially down to values close to zero \cite{Blanter2004, Melnikov2006}.

The introduction of normal-metal layers in such an FNFN...FN composite structure has two major consequences. First, these layers act as buffers that suppress interdiffusion between adjacent ferromagnetic layers. Second, they magnetically decouple the ferromagnetic sublayers, enabling independent control of the magnetization direction $\bm{M}$ in different parts of the multilayer stack. This allows for tunable control of the decay length $\xi_{1}$.

As previously noted, the proximity effect in an FN bilayer leads to renormalization of the effective exchange field $H_{\mathrm{eff}}$ that determines $\xi_{1}$. Physically, this renormalization originates from the fact that an electron can partially reside in the N-layer, where no spin-splitting occurs. This effectively reduces the average exchange energy felt by the Cooper pairs, resulting in a longer decay length for superconducting correlations.

\section{Spin Valves for Cryogenic Memory} 

The concept of a Josephson junction with a controllable critical current via relative magnetic orientation was first proposed theoretically using a ferromagnetic spin-valve structure \cite{oh1997}. Two decades later, this concept was realized experimentally in a hybrid device demonstrating fully controllable 0 to $\pi$ transitions \cite{gingrich2016controllable}, establishing the foundation for cryogenic memory applications.

\begin{figure*}[t]
	\centering
\includegraphics[width=1.99\columnwidth]{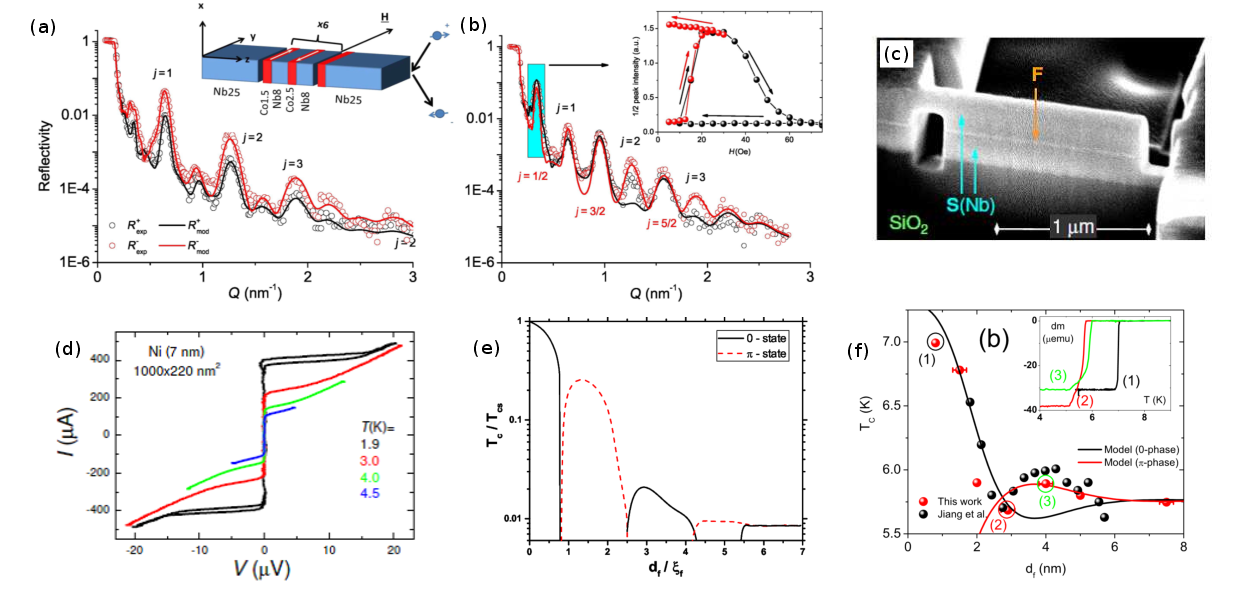}
	\caption{(a, b) Experimental (dots) and simulated (solid lines) specular neutron reflectivity curves measured at $T = 13$ K in applied magnetic fields of $H = 300$ Oe (a) and $H = 30$ Oe (b). 
Insets show: (a) schematic of the sample and PNR setup; (b) magnetic field dependence of the $j = 1/2$ Bragg peak (highlighted in blue). The sample consists of a periodic [Co(2 nm)/Nb(8 nm)]$\times$12 spin-valve structure \cite{Klenov2019}. Numbers above the peaks denote the Bragg reflection order. 
(c) Scanning electron microscope (SEM) image of an SFS junction fabricated using focused ion beam (FIB) etching.  
(d) $I$–$V$ characteristics of a Nb/Ni (7 nm)/Nb junction at various temperatures $T$, illustrating superconducting the Josephson effect \cite{Kapran2021}. 
(e) Calculated dependence of critical temperature $T_c$ on ferromagnetic layer thickness $d_F$; $T_{cs}$ is the intrinsic $T_c$ of the superconducting layer in isolation \cite{Karabassov2019}.  
(f) Experimental $T_c(d_F)$ data from Ref.~\cite{Khaydukov2018} (red dots) and Ref.~\cite{Jiang1995} (black dots). Solid lines are theoretical fits based on Usadel equations for both $0$ and $\pi$ phase states \cite{Karabassov2019} in S/F/S structures using material parameters from the measurements \cite{Khaydukov2018,Jiang1995}. 
Inset shows the temperature dependence of the magnetic moment for the set of samples marked in the main panel (from Ref.~\cite{Khaydukov2018}).
Figures 1 (a,b) were reproduced from \cite{Klenov2019}  (© 2019 N. Klenov et al., published by the Beilstein-
Institut, distributed under the terms of the Creative Commons Attribution 4.0 International License, https://creativecommons.org/licenses/by/4.0). Figures 1 (c,d) were reproduced from \cite{Kapran2021} (© 2021 O. M. Kapran et al., 
published by the American Physical Society, distributed under the terms of the Creative Commons Attribution
4.0 International License, https://creativecommons.org/licenses/by/4.0).
Figure 1 (e) was reprinted with permission from Ref.~\cite{Karabassov2019}, Copyright 2019 by the American Physical Society. This content is not subject to CC BY 4.0. Figure 1 (f) was reprinted with permission from Ref.~\cite{Khaydukov2018}, Copyright 2018 by the American Physical
Society. This content is not subject to CC BY 4.0.
}
	\label{fig_AS}
\end{figure*}

Magnetic structures with tunable effective exchange energy have been successfully developed and investigated for use in Josephson switches and cryogenic superconducting memory elements. One notable implementation is the periodic Co/Nb pseudo-spin-valve structure, as reported in Ref.\cite{Klenov2019}. In this design, the weak-link region of an SFS Josephson junction is formed by a multilayered stack composed of ferromagnetic cobalt (Co) layers separated by thin superconducting niobium (Nb) spacers, as illustrated in the inset of Fig.\ref{fig_AS}(a).

Experimental characterization using polarized neutron reflectometry (PNR), shown in Figs.~\ref{fig_AS}(a,b), revealed a transition from collinear to non-collinear magnetization alignment in the Co layers under relatively low applied magnetic fields $\sim 30$ Oe. This magnetic reconfiguration allows for in-situ control of the effective exchange field, enabling active modulation of the Josephson coupling.

A distinct category of the spin valves is based on the spin-trigger effect\cite{bakurskii2024josephson,Neilo}. This effect is associated with a sharp transition of a thin superconducting layer s from the normal to the superconducting state due to the controlled proximity effect with a bulk superconductor S. In this case, the penetration ability of Cooper pairs can be regulated using a multilayer ferromagnetic structure. The use of this effect allows for the creation of effective spin valves for the development of the elements with tunable inductance \cite{schegolev2022tunable,2025magnetic}.


Advancements in the fabrication of SFS junctions, achieved using focused ion beam (FIB) nanofabrication techniques \cite{bell2003}, enabled the clear demonstration of 0-$\pi$ oscillations in Nb-Permalloy junctions \cite{bell2006}. This fabrication approach was further implemented by Kapran et al. \cite{Kapran2021}, where FIB-etched structures employing ferromagnetic weak links exhibited clear and reproducible switching behavior (see the switching characteristics in Fig.\ref{fig_AS}(c), (d)). These junctions are particularly promising for integration into superconducting memory elements, offering robust switching characteristics and compatibility with scalable cryogenic logic platforms. A detailed study of such junctions using Ni-based ferromagnetic alloys demonstrated a controlled crossover between short- and long-range proximity effects, underscoring the tunability of their superconducting response through magnetic design \cite{Kapran2021}.

Hybrid S/F structures are now widely implemented in the design of novel spintronic devices and circuits\cite{Sidorenko2023}.

\section{Oscillations of $T_c$}

The oscillation of the Cooper-pair wave function within the ferromagnetic (F) layer gives rise to the non-monotonic proximity effect in S/F systems. Specifically, the superconducting critical temperature $T_c$ exhibits oscillatory behavior as a function of the F-layer thickness $d_F$, 
due to constructive and destructive interference of the pairing amplitude \cite{Tagirov1998,Tagirov,Khusainov1997,Fominov2002}. The oscillation's amplitude is highly sensitive to the thickness of the superconducting S-layer. For sufficiently thick S-layers $d_S >> \xi_S$, the outer surface becomes insensitive to perturbations from the F-layer interface.

These mesoscopic oscillations were experimentally demonstrated in a series of Nb/Cu$_{1–x}$Ni$_x$ 
bilayers with variable alloy thickness. Zdravkov et al. (2006) observed clear oscillations in 
$T_c$ versus $d_F$  Nb/Cu in  Nb/Cu$_{0.59}$Ni$_{0.41}$  alloy using detailed structural control \cite{Zdravkov2006}. 
 
\begin{figure}[t]
	\centering
	\includegraphics[width=\columnwidth]{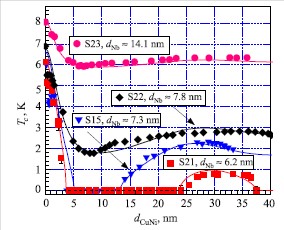}
    \caption{The transition temperature $T_c$ for a series of 
    Nb/Cu$_{0.59}$Ni$_{0.41}$ bilayers. The thickness of the flat Nb layer was fixed in different sample series: $d_{\mathrm{Nb}} \approx 14.1$~nm (S23), $7.8$~nm, $7.3$~nm, and $6.2$~nm. For $d_{\mathrm{Nb}} \approx 6.2$~nm, the transition temperature exhibits fully reentrant behavior. Reproduced from Ref.~\cite{Zdravkov2010}.} 
    	\label{reentrance}
\end{figure}

Not only oscillatory, but also pronounced reentrant superconductivity, where superconductivity is first suppressed and then recovered as a function of 
$d_F$, has been experimentally observed. This is a consequence of the presence of a quasi-one-dimensional FFLO like state in these bilayers. In  \cite{Sidorenko2009,Zdravkov2010} a clear reentrant dependence of the critical temperature $T_c$ was reported, as shown in Fig.~\ref{reentrance}. Further theoretical and experimental modeling by Zdravkov et al. \cite{Zdravkov2011} confirmed the role of interference effects in the pairing wave function in determining superconducting behavior in S/F bilayers \cite{Zdravkov2011,Khaydukov2015,Lenk2016}. Similar effect was observed in Nb/Cu$_{1-x}$Ni$_x$/Nb  trilayers by Kehrle et al. \cite{Kehrle2012}, where the FFLO-like state in the ferromagnetic spacer led to both oscillations and reentrance in $T_c$. Further studies on S/F hybrids and trilayers also confirmed this phenomenon, emphasizing its robustness across various multilayer structures \cite{Zdravkov2012}.

These phenomena became observable thanks to advanced deposition techniques, such as those developed by Morari \textit{ et al.} \cite{morari2012nanolayers}, which significantly improved the quality of ultra-thin Nb films and Nb/CuNi nanostructures compared to earlier fabrication methods.

Figure~\ref{reentrance} demonstrates the dependence of the superconducting transition temperature  on the Cu$_{0.59}$Ni$_{0.41}$ layer thickness for a series of Nb/CuNi bilayers \cite{Zdravkov2010}. It is demonstrated that for $d_{\mathrm{Nb}} \approx 14.1$~nm, the transition temperature exhibits a nonmonotonic behavior with a shallow minimum around $d_{\mathrm{CuNi}} \approx 7.0$~nm. 
In the sample with $d_{\mathrm{Nb}} \approx 7.8$~nm, a more pronounced suppression of $T_c$ is observed, followed by an increase to above 2.5~K, signaling the onset of a reentrant regime. Upon further reduction of the Nb thickness to $d_{\mathrm{Nb}} \approx 7.3$~nm, full reentrant superconductivity emerges: $T_c$ initially decreases with increasing $d_{\mathrm{CuNi}}$ until superconductivity vanishes completely, then reappears at larger $d_{\mathrm{CuNi}}$, reaching values as high as 2~K.

In the thinnest Nb film ($d_{\mathrm{Nb}} \approx 6.2$~nm), $T_c$ drops sharply as $d_{\mathrm{CuNi}}$ increases, and superconductivity is fully suppressed at $d_{\mathrm{CuNi}} \approx 2.5$~nm. In the thickness range $2.5~\mathrm{nm} \leq d_{\mathrm{CuNi}} \leq 24$~nm, no superconducting transition was detected (i.e., $T_c$ falls below the cryostat base temperature of 40~mK). Upon further increase of $d_{\mathrm{CuNi}}$, superconductivity re-emerges at $d_{\mathrm{CuNi}} \approx 25.5$~nm, reaching $T_c \approx 0.8$~K at $d_{\mathrm{CuNi}} \approx 30$~nm, before vanishing again for $d_{\mathrm{CuNi}} \geq 37.5$~nm. 
This striking phenomenon of full suppression and reappearance of superconductivity constitutes the first experimental observation of multiple reentrant behavior in S/F bilayer systems~\cite{Zdravkov2010}.

In S/F/S junctions, the superconducting transition temperature $T_c$ is significantly influenced by the presence of the additional superconducting (S) layer. The Cooper pair wave function in such structures can adopt either a symmetric or antisymmetric configuration relative to the center of the ferromagnetic (F) layer. The symmetry is determined by the structure’s parameters and corresponds to the configuration that minimizes the system's free energy, thereby maximizing $T_c$. This behavior leads to the possibility of $0$–$\pi$ phase transitions, resulting in oscillatory $T_c(d_F)$ dependencies in hybrid S/F/S structures. Theoretical analysis within the framework of quasiclassical Usadel equations confirms that $T_c$ exhibits damped oscillations as a function of the F-layer thickness $d_F$ (see Fig.~\ref{fig_AS}(e)), a result that has been corroborated by multiple experiments (see Fig.~\ref{fig_AS}(f)).

Moreover, under suitable conditions, multiple $0$–$\pi$ transitions can occur, giving rise to complex $T_c(d_F)$ profiles, as predicted theoretically in Ref.~\cite{Karabassov2019}. These phenomena illustrate the sensitive interplay between superconductivity and ferromagnetism in proximity-coupled systems.

\section{Josephson Effect in Junctions with Ferromagnetic Barriers}

The most prominent feature of a CPR of the SFS junction is the possibility
to realize the $\pi $ state. Below we will discuss several mechanisms of 0-$%
\pi $ transitions and their quantitative criteria in terms of $S$, $F$
material parameters and the SF interface resistance. Further, under certain
conditions new types of nontrivial CPR can be realized, having two energy
minima for $\varphi =0$ and $\varphi =\pi $.

\subsection{ 0-$\protect\pi $ Transitions in SFS Junctions}

For an SFS junction with a metallic ferromagnet the possibility of realization of a $\pi $ state was predicted by Buzdin \textit{et al.} in \cite{ISI:A1982PC10900006,ISI:A1992HM98200020} for the clean limit and by Buzdin and Kupriyanov  
\cite{ISI:A1991FT22600011} for the dirty limit. In the latter case, analytical solutions were obtained at temperatures close to $T_C$, which clearly demonstrated oscillations of the critical current as a function of the F-layer thickness $d_{F}$. In the regime, when exchange energy $H$ is large compared to $k_B T_C$, the CPR is sinusoidal with the
critical current \cite{ISI:A1991FT22600011} 

\begin{eqnarray}
I_{C}R_{N} &=&\frac{\pi \Delta ^{2}}{4eT_{C}}y\frac{\sinh y\cos y+\cosh
y\sin y}{\sinh ^{2}y\cos ^{2}y+\cosh ^{2}y\sin ^{2}y},  \label{SFS_GL} \\
y &=&\frac{d_{F}}{\xi _{F}}\sqrt{\frac{H}{2\pi T_{C}}}.  \notag
\end{eqnarray}%
Here $R_{N}$ is the resistance of
the junction, $\xi_F = \sqrt{D_F/2 \pi T_C}$ and $\Delta $ is the value of the pair potential in a
superconductor. Equation (\ref{SFS_GL}) describes damped
oscillations of the critical current as a function of $d_{F}$, where the
negative values of $I_{C}$ correspond to a $\pi $ junction. At large
thicknesses critical current decays as $I_{C}\propto \exp (-d_{F}/\xi _{F1})$%
, while the oscillation period is given by $2\pi \xi _{F2}$, with $\xi
_{F1,2}=\sqrt{D_F/H}$. The
critical current vanishes at $y\approx 3\pi /4+\pi n$.

At arbitrary $T$ simple analytical result can be obtained in the limit of large
thicknesses $y\gg 1$ and under rigid boundary conditions. The resulting CPR is sinusoidal
with critical current given by \cite{ISI:A1982PC10900006,ISI:A1992HM98200020} 
\begin{equation}
I_{C}R_{N}=32\sqrt{2}\frac{\Delta }{e}\mathcal{F}(\Delta /T)y\exp (-y)\sin
(y+\pi /4),  \label{SFS_T_rigid}
\end{equation}%
where $\mathcal{F}(\Delta /T)$ is monotonous function of temperature, $%
\mathcal{F}(\Delta /T)=(\pi /128)\Delta /T_{C}$ at $T\approx T_{C}$ and $%
\mathcal{F}(\Delta /T)\approx 0.071$ at $T\ll T_{C}$. Near transition
temperature the expression (\ref{SFS_T_rigid}) coincides with Eq.~(\ref%
{SFS_GL}). Thus, the damped oscillatory behavior of $I_{C}$ vs $d_{F}$ holds
in whole temperature range.

Let us now consider the case of a weak ferromagnet, when thermal energy $k_BT$
is not negligible compared to exchange energy $H$. 
If interface transparency is low ($\gamma _{B}\gg 1$), the Usadel equations
in the F layer can be linearized and the supercurrent is given by \cite{1128994620010312}
\begin{equation}
I_{S}=\frac{4\pi T}{eR_{N}}\frac{d_{F}}{\gamma _{B}\xi _{F}} Re
\sum\limits_{\omega >0}\frac{\Delta ^{2} \sin(\varphi)}{(\omega ^{2}+\Delta
^{2}) \widetilde{d}_{F}\sinh \widetilde{d}_{F}},  \label{Ryaz}
\end{equation}%
where $\widetilde{d}_{F}=d_{F}/\xi $ and $\xi$ is given by Eq.~(\ref{xi}).
In this regime the temperature dependence of $I_{C}$ is
not sensitive to the value of $\gamma _{B}$, while the magnitude of $%
I_{C}R_{N}$ product is suppressed as this parameter rises. It also follows
from Eq.~(\ref{Ryaz}) that $I_{C}$ oscillates vs $d_{F}$ with the
temperature-dependent period $\xi _{F2}$ given by Eq.~(\ref{xi}). This
provides the possibility of 0-$\pi $ crossover as temperature decreases, if $%
H\thicksim \pi T_{C}$ and $d_{F}\thicksim \xi _{F2}$.
Such a temperature driven crossover was observed by Ryazanov \textit{et al. }
\cite{1128994620010312} in Nb/Cu$%
_{1-x}$Ni$_{x}$/Nb Josephson junctions.

Theoretical models for tunnel Josephson junctions with ferromagnetic interlayers have been refined to account for non-sinusoidal CPRs and higher harmonic contributions. Vasenko et al. \cite{Vasenko2008,Vasenko2011} systematically analyzed the impact of ferromagnetic layer thickness and interface transparency on the current-phase relation, showing that the junction’s critical current and ground state phase can be tuned by varying these parameters. These studies bridge the gap between idealized models and realistic experimental conditions, particularly for junctions incorporating weak ferromagnets like CuNi alloys.

Further studies have explored the anomalous transport properties of SFIFS Josephson junctions with weak ferromagnetic interlayers. Karabassov et al. \cite{Karabassov2020,Karabassov2022} reported non-monotonic current-voltage characteristics in such structures, attributed to the interplay between spin-singlet and spin-triplet pairing correlations. These findings complement the understanding of 0-$\pi$ transitions in SFS-type junctions and highlight the role of weak ferromagnets in tailoring Josephson coupling.

In a series of studies, Birge and coauthors \cite{mishra2023effect,mishra2022enhancement,aguilar2020spin,birge2019spin,PhysRevB.99.174519,PhysRevB.97.214509} experimentally realized spin-triplet supercurrents in SFS junctions with various types of ferromagnetic barriers using a number of material combinations. They also discussed their prospects for applications as circuit elements in superconducting logic and memory. 

A recent review \cite{ryazanov2025josephson} summarizes the features of the transition to the $\pi$ state in SFS junctions and the nonequilibrium distributions arising in hybrid Josephson S-N/F-S structures upon injection of spin-polarized carriers into the barrier.

\subsection{ 0-$\protect\pi $ Transitions in SIsFS Junctions}

The emerging prospects of practical applications for Josephson junctions with magnetic barriers have stimulated interest in their use as elements of superconducting memory devices~\cite{Larkin,Vernik,Nevirkovets1,Nevirkovets2,Bakurskiy1,Bakurskiy2}, on-chip $\pi$-phase shifters for self-biasing various electronic quantum and classical circuits~\cite{1001526820080515}, and $\varphi$-batteries—structures with a nonzero phase difference $\varphi$ in their ground state~\cite{Buzdin_fi,Pugach_fi,bakurskii2013josephson3130659,Goldobin2013,Gold2,Heim,Linder1,Chan_HM} and led to the development of a new class of SFS-based devices, namely, SIsFS junctions.
Recent work exploring the use of ferromagnetic Josephson junctions for cryogenic memory includes \cite{Caruso,Shafraniuk, Nevirkovets3, parlato2020characterization}. These applications have driven the development of a new class of SFS devices, namely, SIsFS junctions.

These structures have been the subject of growing theoretical and experimental research. Experimentally, scalable memory elements based on rectangular SIsFS geometries have been demonstrated and evaluated for practical cryogenic memory applications \cite{Karelina2021}. Newer designs even explore their potential as tunable phase elements for superconducting qubits and quantum hybrid systems 
\cite{satariano2024nanoscale,ahmad2024phase}, further expanding the functionality of SIsFS junctions in superconducting circuit architectures.

The SIsFS structure represents a series connection of an SIs tunnel junction and an sFS contact. The properties of SIsFS junctions are governed by the thickness of the intermediate superconducting layer $d_s$ and by the relationship between the critical currents $J_{C,\mathrm{SIs}}$ and $J_{C,\mathrm{sFS}}$ of the SIs and sFS sections, respectively \cite{Bakurskiy1,Bakurskiy2, bakurskiy2017current, bakurskiy2018protected, bakurskiy2021density, Neilo}. When the thickness $d_s$ of the s-layer significantly exceeds its superconducting coherence length $\xi_S$ and the critical current of the SIs junction is much smaller than that of the sFS part ($J_{C,\mathrm{SIs}} \ll J_{C,\mathrm{sFS}}$), the characteristic voltage of the entire SIsFS device is predominantly determined by the SIs segment and may approach the maximum value typical of standard SIS junctions. At the same time, the phase difference $\varphi$ in the ground state of the SIsFS junction is dictated by the sFS region, allowing the system to reside in either the 0- or $\pi$-state depending on the thickness of the ferromagnetic layer. 
This tunability opens a pathway to implementing controllable $\pi$-junctions that maintain a large $J_C R_N$ product, crucial for applications in superconducting memory and cryogenic digital circuits \cite{Nevirkovets2, satariano2024nanoscale}.

\begin{figure*}[tbp]
\center{\includegraphics[width=12 cm]{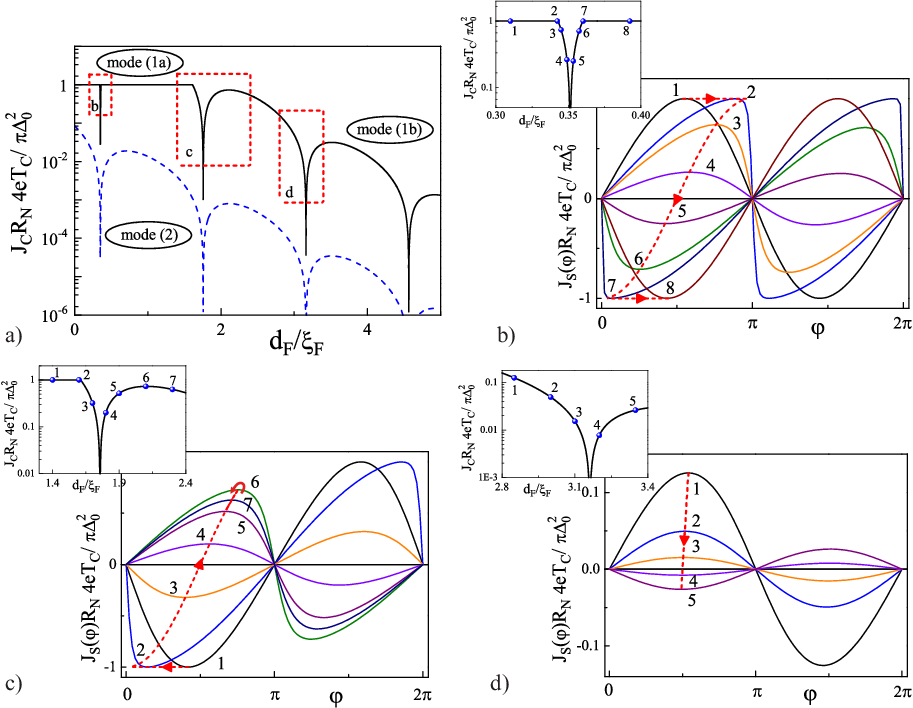}}
\caption{(a) Dependence of the critical current density $J_C$ of the SIsFS junction on the F-layer thickness $d_F$ for two values of the intermediate s-layer thickness: $d_s = 5\xi_S(T) > d_{sc}$ (solid line) and $d_s = 0.5\xi_S(T) < d_{sc}$ (dashed line). The calculations are performed at $T = 0.9T_C$ for $H = 10\pi T_C$ and $\Gamma = 5$. 
(b–d) Current-phase relations $J_S(\varphi)$ in the vicinity of $0$–$\pi$ transitions. Each panel includes an inset that enlarges a corresponding section of the $J_C(d_F)$ curve from part (a), marked by the letters b–d. The numbered points in the insets indicate the specific $d_F$ values at which the $J_S(\varphi)$ curves were computed. Dashed lines in panels (b)–(d) trace the critical points where $J_S(\varphi)$ attains its maximum value $J_C(d_F)$. 
Reprinted with permission from Ref.~\cite{Bakurskiy2}, Copyright 2013 by the American Physical Society. This content is not subject to CC BY 4.0.}
\label{CPR}
\end{figure*}

The behavior of SIsFS junctions was analyzed self-consistently in Ref.~\cite{Bakurskiy2} and the results are presented in Fig.\ref{CPR}(a-d). Figure\ref{CPR}a shows the calculated dependence of the characteristic voltage $R_N J_C(d_F)$ as a function of the ferromagnetic layer thickness $d_F$, evaluated at $T = 0.9T_C$ for representative material parameters: exchange field $H = 10\pi T_C$, ideal transparency at the sF interface ($\gamma_B = 0$) and interface transparency parameter $\Gamma \approx 5$.

Two cases are considered for the thickness of the intermediate superconducting s-layer: $d_s = 5\xi_S(T)$ (solid line) and $d_s = 0.5\xi_S(T)$ (dashed line), where the temperature-dependent coherence length is given by
\begin{equation}
\xi_S(T) = \frac{\pi \xi_S}{2\sqrt{1 - T/T_C}}, \quad \Gamma = \frac{\gamma \xi_S(T)}{\xi_S}.
\label{GammasB1}
\end{equation}
Here, $\gamma$ and $\gamma_B$ denote the suppression parameters of the sF and SF interfaces, respectively.

Figures~\ref{CPR}b–d present magnified views of the regions marked by rectangles in Fig.~\ref{CPR}a. Each subfigure focuses on a narrow $d_F$ interval exhibiting rich behavior of the current-phase relation (CPR) $J_S(\varphi)$. The numbered points along the $R_N J_C(d_F)$ curves identify specific $d_F$ values where the CPR was explicitly calculated; the corresponding $J_S(\varphi)$ curves are labeled with the same numbers. Dashed lines in these panels trace the critical points—values of $d_F$ at which $J_S(\varphi)$ reaches its maximum, i.e., the critical current $J_C(d_F)$. These detailed calculations illustrate how subtle variations in $d_F$ near 0–$\pi$ transition boundaries can give rise to strongly anharmonic CPRs and highlight regimes where $\varphi$-junction behavior may emerge.

Figure~\ref{CPR}b illustrates the transition from the $0$-state to the $\pi$-state in the SIsFS structure. Within this narrow interval of $d_F$, the SIsFS junction behaves effectively as a series combination of an SIs tunnel junction and an sFS Josephson contact. The total critical current of the device is then determined by the weaker link, i.e., the smaller of the two critical currents: $J_{C,\mathrm{SIs}}$ or $J_{C,\mathrm{sFS}}$.
In the present case, the s-layer thickness is sufficiently large to preserve its superconducting properties and maintain a nearly constant $J_{C,\mathrm{SIs}}$ as $d_F$ is varied. Thus, in the progression from point 1 to point 2 along the $R_N J_C(d_F)$ curve, $J_C$ is limited by the sFS part. At point 2, the critical currents of the two sections become equal, $J_{C,\mathrm{SIs}} = J_{C,\mathrm{sFS}}$. This point corresponds to the maximum distortion of the current-phase relation (CPR) $J_S(\varphi)$ from the conventional sinusoidal form, a hallmark of the strong interference between the two junction components.
As $d_F$ is increased beyond point 2, the structure undergoes a $0$–$\pi$ transition. The CPR regains a nearly sinusoidal shape immediately after the transition, albeit with an inverted sign of the critical current, reflecting the $\pi$-state of the junction. Continuing from point 6 to point 7, the CPR becomes increasingly anharmonic again, reaching a secondary maximum in distortion at point 7. Further increase in $d_F$ toward point 8 results in the restoration of a sinusoidal CPR, indicating a stabilization of the $\pi$-state regime with minimal higher harmonic content.

Figure~\ref{CPR}c shows the evolution of the current-phase relation (CPR) during the transition from the $\pi$-state back to the $0$-state as the ferromagnetic layer thickness $d_F$ is further increased. The transformation of $J_S(\varphi)$ from point 1 to points 2–5 along the $R_N J_C(d_F)$ curve closely mirrors the behavior observed in Fig.~\ref{CPR}b for the $0$–$\pi$ transition, except for the initial negative sign of the critical current, characteristic of the $\pi$-state. As in the previous case, the CPR undergoes progressive anharmonic deformation, reaching maximal deviation at intermediate points before returning to a sinusoidal form.
This return to the $0$-state manifests as a pronounced kink in the $R_N J_C(d_F)$ dependence, indicating a nonmonotonic variation of the critical current through the transition. At point 6, the critical current reaches a local maximum, after which it begins to decrease along the dashed trajectory as $d_F$ continues to increase, signaling the dominance of the sFS part in limiting the overall current.
Figure~\ref{CPR}d illustrates the behavior of the CPR in the vicinity of a subsequent $0$–$\pi$ transition. At point 1, a slight deviation from the sinusoidal shape is observed. This distortion rapidly vanishes with increasing $d_F$, indicating exponential suppression of higher harmonics and a swift restoration of the sinusoidal CPR in the deeper $\pi$-state regime.

For a small thickness of the s-layer (as shown by the dashed curve in Fig.\ref{CPR}a), intrinsic superconductivity within the s-film is entirely suppressed. This leads to the formation of a composite weak link of the type SInFS, and the resulting current-phase relation (CPR) becomes purely sinusoidal.

The deviations of the $R_N J_S(\varphi)$ dependencies in Fig.~\ref{CPR} from the ideal $\sin(\varphi)$ shape are attributable to the presence of higher harmonics in the CPR. However, in all cases considered here, the amplitudes of these harmonics remain small in comparison to that of the fundamental harmonic. Consequently, the $R_N J_S(\varphi)$ relations remain single-valued functions of $\varphi$. This behavior is characteristic of conventional Josephson devices, such as SNS, SFS, and SINIS junctions, and does not depend significantly on the geometry or specific transport mechanisms within the weak link.

In contrast, when the weak-link region is composed of a material that is intrinsically superconducting (denoted s), with a critical temperature $T_{cs}$ lower than that of the adjacent S electrodes, a qualitatively different behavior may emerge. In such systems, increasing the distance between the electrodes can lead to a transition from a single-valued to a multi-valued $R_N J_S(\varphi)$ dependence \cite{LikYak}. This transition marks a crossover from the Josephson regime to a state dominated by Abrikosov vortex flux flow in the s-layer \cite{KLM}, reflecting the onset of phase slip or vortex dynamics within the superconducting weak link.

\begin{figure}[tbp]
\center{\includegraphics[width=\columnwidth]{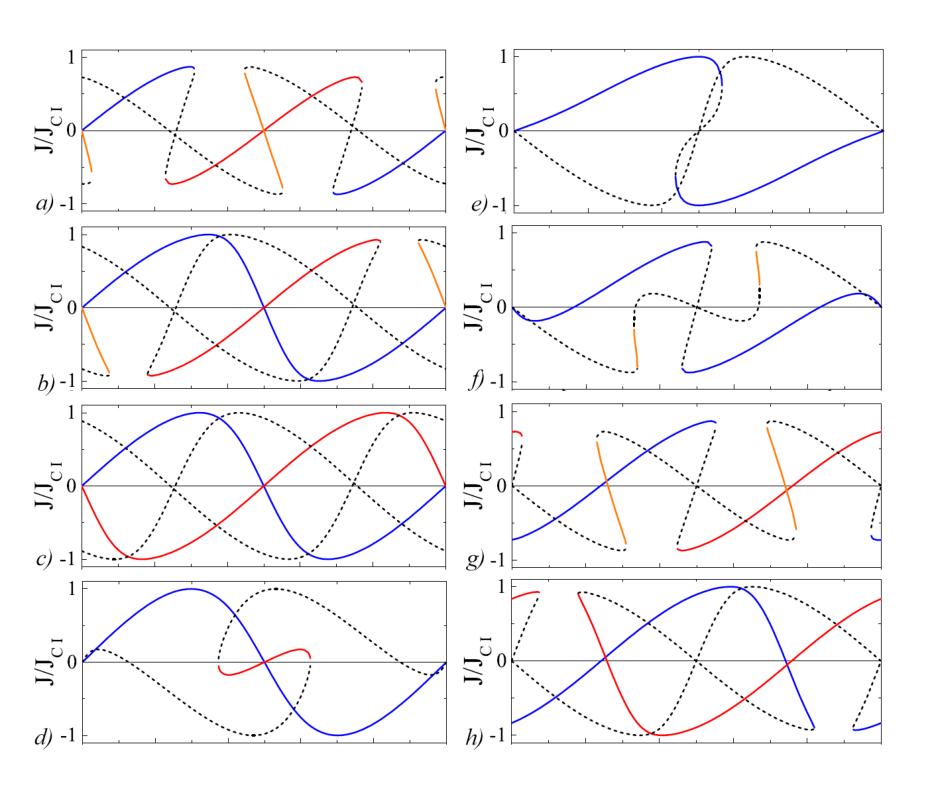}}
\caption{Multi-valued  $R_{N}J_{S}(\varphi )$ dependencies in an SIsFS junction for different values of the first (A) and second (B) harmonics of the CPR of the sFS junction: a) A=0.1, B=0.8; b) A=0.1, B=1.0; c) A=0.1, B=1.2; d) A=1.4, B=1.0; e) A=1.0, B = -0.3; f) A=0.5, B = -0.5; g) A=0.1, B = -0.8; g) A=0.1, B = -1.0. Solid and dashed lines show stable and unstable states, respectively.}
\label{Jphi}
\end{figure}

An SIsFS junction provides an example of the structure with multi-valued  $R_{N}J_{S}(\varphi )$ dependence. In
recent studies  \cite{Bakurskiy2,bakurskiy2017current} it was demonstrated that 
at low $T$   the existence  of significant second harmonic
in current-phase relation (CPR) in sFS part of the structure results in 
developing of instabilities near $0$-$%
\pi$ transition providing  transition to  multi-valued CPR. 
The transition goes through distinct states with a discontinuous hysteretic CPR. 
As demonstrated in Ref.~\cite{Bakurskiy2}, a trivial state within the $0$–$\pi$ transition region is realized when the second harmonic of the current-phase relation (CPR) is small. In this regime, the system supports two degenerate, stable ground states at $\varphi = 0$ and $\varphi = \pi$, both lying on a single, continuous CPR branch. Similar features were also reported in the context of $0$–$\pi$ transitions in conventional SFS junctions \cite{Radovic1, Radovic2}.

The evolution of the CPR in an SIsFS junction with increasing second-harmonic amplitude is shown in Figure~\ref{Jphi} (based on Ref.~\cite{bakurskiy2017current}). A dominant second harmonic induces a qualitatively different, strongly anharmonic CPR, characterized by a multivalued structure. This results in bistability and hysteresis, where sweeping the phase difference $0 \leq \varphi \leq 2\pi$ causes two distinct jumps between stable CPR branches, marking transitions between the system's coexisting ground states.

In this nonlinear regime, additional metastable branches may also emerge at higher energy levels. Although mathematically valid, these states do not represent thermodynamic minima and are inaccessible via adiabatic phase evolution due to large energy barriers separating them from the ground states \cite{Bakurskiy2}.

A particularly interesting regime arises when the weak link is located at the SIs interface, and the SIsFS structure can exist in either the $0$- or $\pi$-state. In this case, the energy–phase relation $E(\varphi)$ features two minima: a global minimum at $\varphi = 0$ and a local minimum at $\varphi = \pi$, or vice versa, depending on the specific parameters of the junction. The energy difference between the two states is small, yet each state is topologically protected in the sense that a transition between them cannot occur via a continuous, adiabatic change of the phase. To switch the system from one state to the other, the bias current must be increased beyond the critical current of the sFS part of the structure.
 
Replacing a single F layer with an F$_1$s$_1$F$_2$ control unit removes the restriction on the thickness of the s film adjacent to the I layer in the SIsFS structure \cite{bakurskii2024josephson}. By exploiting the trigger effect in the SF$_1$s$_1$F$_2$s electrode \cite{Neilo} of this SF$_1$s$_1$F$_2$sIS spin valve, one can switch the junction between its $0$ and $\pi$ states. For an s-layer thickness in the range of 2--3 $\xi_s$, the magnetic subsystem of the spin valve controls only the sign of the order parameter in the s-layer (i.e., whether the junction is in the $0$ or $\pi$ state). The magnitudes of the characteristic voltage in these two states are equal to each other and are close to the value for standard SIs tunnel junctions.

Furthermore, the $0$–$\pi$ transition in this configuration may occur without a visible change in the critical current, instead manifesting through a transformation of the CPR. This complicates conventional methods of detection and necessitates the use of phase-sensitive experimental techniques, as discussed in Ref.~\cite{Frolov,Bauer_2004}. 

\section{Summary}

This review covers recent progress in the physics and applications of superconductor–ferromagnet hybrid structures. We begin with the fundamentals of the proximity effect in SF bilayers and SFS Josephson junctions, where superconducting correlations penetrate the ferromagnet and undergo spatial oscillations due to the exchange field. These oscillations lead to the formation of $\pi$  states, as well as oscillations and reentrant behavior of the critical temperature. Consequently, Josephson junctions containing ferromagnetic layers—such as SFS and SIsFS junctions—exhibit unique current–phase relations. These can include non-sinusoidal and hysteretic behavior, which originate from higher harmonic contributions.

In SIsFS junctions, the interplay between the SIs and sFS parts allows for control over the ground state phase, enabling either $0$ or $\pi$ junction configurations. These structures can support bistable CPRs, making them suitable for use as memory elements. Their transport properties depend crucially on the thickness of the intermediate s-layer. When this superconducting s-layer is thin, the system behaves as a complex SFS-type weak link with a purely sinusoidal CPR. For thicker s-layers, the critical current is determined by the limiting segment (either SIs or sFS), and $0$-$\pi$ transitions can occur without changes in the critical current magnitude

The review also highlights the ability to tune the effective exchange energy $H_{\text{eff}}$ in multilayered pseudo-spin valves (e.g., Co/Nb superlattices), where neutron reflectometry shows magnetic switching at low fields. These engineered materials allow precise control of the superconducting state through magnetic configuration.

Experimental studies on Nb/CuNi bilayers and trilayers demonstrate nonmonotonic and even reentrant $T_c(d_F)$ behavior. These effects are attributed to interference of the superconducting wavefunction inside the ferromagnet and are described using the Usadel equations. Multiple $0$-$\pi$ transitions and FFLO-like modulated states are supported by both theory and experiment.

Overall, SF structures provide a highly tunable platform for advancing superconducting devices, including cryogenic memory, $\pi$-junctions, $\varphi$-batteries, and logic elements. The reviewed works combine theoretical modeling with experimental confirmation, laying the groundwork for practical SF-based quantum electronics.

\section{ACKNOWLEDGEMENTS}

The authors are thankful to V.V. Ryazanov, A.I. Buzdin, Ya. V. Fominov, I. Soloviev, N. Klenov, L. Tagirov, E. Il'ichev and R. Tidecks for numerous discussions.

\section{FUNDING}

\noindent The section ``Proximity Effect'' was prepared under support by the Russian Science Foundation (Grant 23-72-30004).\\
The section ``Spin Valves for Cryogenic Memory'' was prepared under support by the Russian Science Foundation (Grant 25-19-00057).\\
The section ``Oscillations of $T_c$'' was prepared under support by the Ministry of Science and Higher Education of the Russian Federation (Grant FSMG-2023-0014).\\
The section ``Josephson Effect in Junctions with Ferromagnetic Barriers'' was prepared under support by the Ministry of Science and Higher Education of the Russian Federation (Grant 075-15-2025-010).\\
The data analysis of various spin valve structures was supported by the Higher School of Economics project ``International Academic Cooperation.''

\section{conflict of interest}
The authors of this work declare that they have no conflicts of interest.

\bibliography{ht_total2}

\end{document}